\documentclass[journal]{./IEEEtran/IEEEtran}

\usepackage{color}

\usepackage{amsmath, amssymb, amsthm}
\usepackage{graphicx, pstricks, pst-node}
\usepackage{ulem}

\begin{document}

\title{Controlled Tripping of Overheated Lines\\ Mitigates Power Outages}

\author{Ren\'{e}~Pfitzner,
        Konstantin~Turitsyn,
        and~Michael~Chertkov% <-this % stops a space

\thanks{
The work at LANL was carried out under the auspices of the
National Nuclear Security Administration of the U.S. Department of
Energy at Los Alamos National Laboratory under Contract No.
DE-AC52-06NA25396.
The work of RP and MC is partially supported by NSF grant DMS-0807592.
The work of MC was funded in part by DTRA/DOD under the grant BRCALL06-Per3-D-2-0022 on ``Network Adaptability from WMD Disruption and Cascading Failures".
}
%\thanks{R.P. is a student at the Chair of Systems Design, ETH Z\"{u}rich, Switzerland as well as at New Mexico Consortium and CNLS, Los Alamos National Laboratory, Los Alamos, NM {\tt\small pfitzner.rene@gmail.com}}
%\thanks{K.T. is with MIT, Department of Mechanical Engineering, {\tt\small turitsyn@mit.edu}}
%\thanks{M.C. is with Theory Division \& Center for Nonlinear Studies at LANL,
%Los Alamos, NM and also with New Mexico Consortium, Los Alamos, NM {\tt\small chertkov@lanl.gov}}
}

%\markboth{submitted to IEEE SmartGridComm 2011}
%{Shell \MakeLowercase{\textit{et al.}}: Bare Demo of IEEEtran.cls for Journals}

\maketitle

\begin{abstract}
We study the evolution of \textit{fast} blackout cascades in the model of the Polish (transmission) power grid ($2700$ nodes and $3504$ transmission lines). The cascade is initiated by a sufficiently severe initial contingency tripping. It propagates via \textit{sequential} trippings of many more overheated lines, islanding loads and generators and eventually arriving at a fixed point with the surviving part of the system being power-flow-balanced and the rest of the system being outaged. Utilizing an improved form of the quasi-static model for cascade propagation introduced in our earlier study (\textit{Statistical Classification of Cascading Failures in Power Grids}, IEEE PES GM 2011), we analyze how the severity of the cascade depends on the order of tripping overheated lines. Our main observation is that \textit{the order of tripping has a tremendous effect on the size of the resulting outage}. Finding the ``best" tripping, defined as causing the least damage, constitutes a difficult dynamical optimization problem, whose solution is most likely computationally infeasible. Instead, here we study performance of a number of natural heuristics, resolving the next switching decision based on the current state of the grid.
Overall, we conclude that \textit{controlled intentional tripping is advantageous in the situation of a fast developing extreme emergency, as it provides significant mitigation of the resulting damage}.
\end{abstract}


\IEEEpeerreviewmaketitle

\section{Introduction}

The effect of large power grid blackouts on the economy and on our everyday life is enormous. Unfortunately the grid of today in the US, and also in many other countries, operates on the edge, thus making large and costly blackouts, such as the August 2003 East Coast blackout, more and more probable. Increase in energy consumption with a pace exceeding reinforcement of the power systems, growing fluctuations  (e.g. associated with intermittency of new renewable sources) and insufficient upgrade of the transmission system are the major factors leading to increase of the failure probability.
This paper contributes to the recent line of research motivated by this growing and important problem.

Cascades are extreme, and hopefully rare, processes.  In this manuscript we aim to analyze the final damage of a cascading process taking place on the time scale of tens of seconds to minutes, and to find a way of controlling the cascade and minimizing its final damage via a carefully selected and automatically executed sequence of line-trippings. The logic here is that controlled tripping of overloaded lines, as a replacement for the ``do nothing" scheme, i.e. ``waiting" for probabilistically natural tripping, might be beneficial in that it redistributes power-flows in a favorable way. Our results suggest that this is indeed a valid assumption. This type of emergency control, however, will require on-the-fly state estimation for computations and eventually selecting optimal (or just good) control actions. The emergency setting also assumes a flawless execution of these control actions. Both, discovering and executing the cascade-mitigating strategy of the line trippings, impose significant constraints on communications. In this regards, our analysis emphasizes the importance of fast and reliable communications necessary to mitigate fast emerging cascades.

\begin{figure}[t]
\includegraphics[width=1\columnwidth]{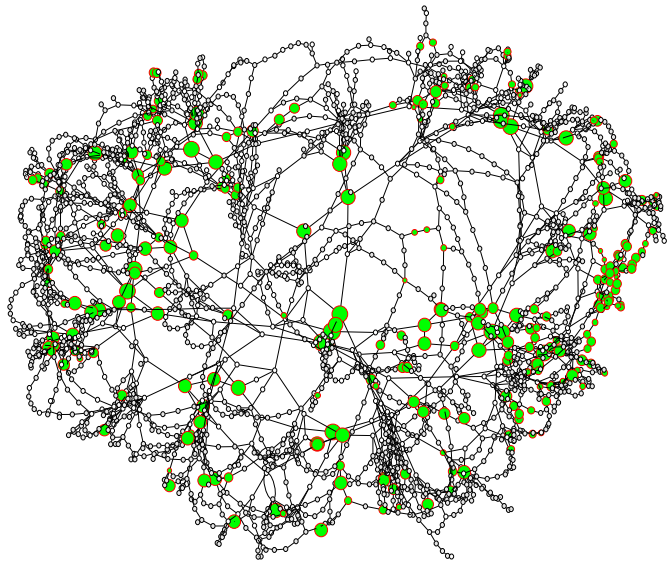}
\caption{Visualization of the Polish power grid (non-geographical). Small grey circles denote consumer nodes. Bigger green circles denote generators. The bigger the (green) circle the larger power generated.}
\label{fig:polish1}
\end{figure}

A number of modeling methodologies were developed to study cascades, see \cite{CarLynDobNew2004, HinBalSan2009, InitRev2008} for comprehensive reviews. Here we choose to work with a framework of microscopic modeling and simulations of cascades, originated from \cite{CheThoDob2005,NedDobKirCarLyn2006} and continued in our recent paper \cite{PfiTurChe2011} \footnote{Note that this framework is different from the computationally advantageous but pure phenomenological modeling of \cite{BarAlb1999,Watts2002,CruLatMar2004}. Once the microscopic approaches,  of the type discussed in this manuscript, are developed, they should  help selecting the right phenomenological candidate proper for quantitative power grid modeling. The general importance of models which go beyond pure graph-theoretical considerations was pointed out and quantified before in \cite{HinCotBlu2010}.}.

We study cascades in the power grid model of Poland, shown in Fig.~(\ref{fig:polish1}) in a non-geographical format, rendered using Graphviz~\cite{Graphviz}. This model is publicly available as a part of the MATPOWER computational package \cite{ZimSanTho2010}. The base case of the studied model corresponds to a peak load (in the summer of 2004) of 18GW, which accounts to $\sim 60\%$ of the total generation capacity. We follow the approach of \cite{PfiTurChe2011, Bienstock2010} and resolve power-flows within the Directed Current (DC) approximation \cite{Kundur1994}. We initialize a cascade by tripping one or two lines in the base case. We propagate the cascade via resolving power-flow equations. Then we trip randomly one of the overloaded lines, check for islands (disconnected components) and
implement mandatory load shedding and/or outage islands containing insufficient generation to supply the load. We use the standard scheme of \textit{droop control} to redistribute generation at any of the steps in the process. We iterate until a balanced solution on the surviving part of the grid emerges. (See Fig.~\ref{fig:flowchart} for the flowchart of the scheme, based on \cite{PfiTurChe2011},  with respective explanations of the scheme briefly repeated in Section \ref{sec:Model}.)
We analyze the dependency of the final damage (measured in terms of the fraction of power demand not served, but also in terms of the number of steps it takes to stop the cascade) on the specific choice of the tripping sequence of overloaded lines.
We pose the question of finding the optimal strategy of line trippings, i.e. the strategy leading to the least final outage, and stress that this question is a difficult one to answer precisely. Thus, we settle here with the analysis of heuristic algorithms proposing to choose the next-to-be-tripped line based on the current status of the power flows. We consider four different heuristics,  test their performance on the Polish model and compare and discuss the results in Sections \ref{sec:Random},\ref{sec:Control}.
A brief description of our main findings is as follows:
\begin{itemize}
\item \underline{Random Tripping}. (See Section \ref{sec:Random} for details.) For any initial tripping we consider multiple memoryless and statistically uniform tripping paths, that is at each instance of time we pick at random one of the overloaded lines to become the next one to be tripped. In general, the resulting distribution of final outage size is surprisingly broad, with some tripping paths being almost ideal (in terms of leading to either no or very small outage) while others resulting in outaging a very significant portion of the grid. This observation suggests that one benefits by not waiting and thus effectively allowing the lines to trip randomly, but instead initiating an optimal tripping, resulting in a minimal finite damage.

\item \underline{Control heuristics}. (See Section \ref{sec:Control} for details.) Finding the optimal switching strategy is a difficult task, which most probably results in a solution which cannot be stated in terms of some graph-local rules. Aiming to reduce complexity of this task, we rely on (and test) four intuitive and simple graph-local heuristics.  We observe, that even these relatively simple heuristics may mitigate a severe cascade. Heuristic, which according to our experiments performs the best in most (but not all) cases, selects at any instance to trip the least loaded of all the overloaded lines. We relate this surprisingly good performance behavior to a (simplified) hierarchical structure of the power grid.
\end{itemize}
We summarize our results and discuss future research directions in Section \ref{sec:Summary}.

\section{Our Model}
\label{sec:Model}

There exist multiple failure mechanisms which can lead to cascades. The most standard failure is an incidental line tripping. When the operational conditions are normal, tripping of a line is a low probability event for example associated with a tree falling on the line.  However, if the operation becomes abnormal and the power flowing through a line exceeds its threshold capacity, the line tripping becomes almost inevitable. In this extreme regime any small external initiation,  for example associated with a modest wind or a perturbation caused by a bird flying near by, will almost certainly (with probability one) result in a short circuit to the ground, and thus inevitably lead to tripping in a matter of minutes. Motivated by these considerations, we assume in our modeling approach that the cascade is initiated by a small number of co-incidental events, leading for example to simultaneous tripping of one or two strong lines.   If this initial failure is sufficiently large,  it leads (after resolving the power-flow equations) to some other lines exceeding their thermal limits.  These overheated lines are not tripped instantaneously, but almost certainly one line will be tripped (due to increased failure probability) within few minutes, if no operator action is taken. It is natural to assume, that in this relatively short time span left for the overloaded situation to be resolved, all other external characteristics, such as configurations of loads and uncontrolled generation (e.g. associated with renewables), remain unchanged. When developed naturally and not mitigated, the cascade may be very large and damaging. This motivates us to focus on possible preventive actions the operator (or an automatic control system) can take during the period when the line already exceeded its rating but it is not yet tripped. As we will see below, it may be advantageous to trip overloaded lines in a special order, minimizing the resulting final outage. Indeed, tripping a line (i.e. changing the underlying topology) redistributes flows and thus,  if the order is chosen wisely, can lead to a better redistribution of the outages over the grid, possibly relieving lines from overload, or at least reducing the remaining overload.

Our model reflects the existing reality of the power grid automatic control. We assume that generation re-dispatch,  which is typically done every 15 minutes to an hour, is not available for fast adjustment necessary to mitigate the emergency overload. Hence, in the here considered timescale, the system continues to operate under the primary (droop) control, also supported by emergency load shedding. See Section \ref{subsec:droop_control} for detailed discussion.

\begin{figure}[t]
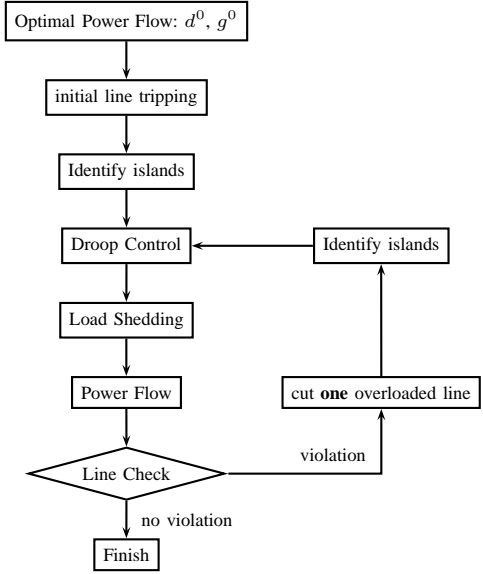

\begin{center}
{\scriptsize
\begin{psmatrix}[colsep=0.05\columnwidth,rowsep=0.5cm]
\rnode{A}{\psframebox{Optimal Power Flow: $d^0$, $g^0$}}\\
\rnode{B}{\psframebox{initial line tripping}}\\
\rnode{B_1}{\psframebox{Identify islands}}\\
\rnode{E}{\psframebox{Droop Control}}&\rnode{E2}{\psframebox{Identify islands}}\\
\rnode{EF1}{\psframebox{Load Shedding}}\\
\rnode{F}{\psframebox{Power Flow}}&\rnode{F2}{\psframebox{cut \textbf{one} overloaded line}}\\
\dianode{G}{Line Check}\\
\rnode{H}{\psframebox{Finish}}
\ncline{->}{A}{B}
\ncline{->}{B}{B_1}
\ncline{->}{B_1}{E}
\ncline{->}{E}{EF1}
\ncline{->}{EF1}{F}
\ncline{->}{F}{G}
\ncline{->}{G}{H}
\naput{no violation}
\ncline{->}{F2}{E2}
\ncline{->}{E2}{E}
\ncangle[angleB=-90]{->}{G}{F2}
\naput{violation}
\end{psmatrix} }
\caption{Flowchart of our microscopic quasi-static model of the cascade.}
\label{fig:flowchart}
\end{center}
\end{figure}

The flowchart diagram shown in Fig.~\ref{fig:flowchart} explains our cascade model. We initiate the model with an Optimal Power Flow (OPF) solution and introduce initial failures by tripping one or more lines. We then check for islanding and include the droop control mechanism to match generation and demand. Next, the scheme evaluates the PF solution \footnote{We choose to work here with the DC approximation, ignoring variations in voltage and resistive characteristics of lines, and assuming that the phase difference across any of the lines is sufficiently small. This approximation is realistic and it also helps us to make the algorithm lighter and focus more on the most important new ingredients of the model, associated with the sequence of trippings and network effects. On the other hand,  our scheme certainly allows running more accurate AC power flow instead of DC,  and thus accounting for voltage constraints at the nodes where voltage is not controlled directly.} on the new model, checks if line flows exceed ratings and, if so, trip exactly \textbf{one} of these lines. Let us emphasize that the choice of tripping a single line at once reflects the physical reality of the grid better than simultaneous tripping of many lines one may consider to simplify the model. Indeed, if the tripping occurs by itself because the line exceeds its thermal limit,  the event is random and it is not correlated to other trippings. Also the typical time between consecutive trippings is significantly longer than the time for all the electric and electro-mechanical transients to occur. (The transients are settled in seconds or even faster.) Additionally,  if the tripping would be initiated by a control system, it is safer to trip lines one-by-one to avoid strong perturbations and stronger transients. The cascade algorithm is repeated until all the thermal constrains are resolved, and a steady feasible solution is achieved. Different stages of the algorithm are described in detail in the following subsections, which mainly follows the logic of \cite{PfiTurChe2011}, also with addition of a new ingredient - mandatory load shedding over an island with an insufficient generation.
\subsection{DC power flow}
\label{subsec:dc_power_flow}

A general power-flow solver takes injection and consumption of powers at all the nodes of the power grid and other system parameters as input and outputs voltages and phases at the nodes and powers transmitted over all the transmission lines of the grid. Our cascading algorithm will work with the most general power solver. However, and as in \cite{PfiTurChe2011}, we choose to work with a DC solver, which is a bit simpler in implementation. The DC solver evaluates
\begin{eqnarray}
&& \forall i\in{\cal G}_0:\quad \sum_{j\sim i}
p_{ij}=\Biggl\{\begin{array}{cc}
 g_i, & i\in{\cal G}_g\\
 -d_i, & i\in{\cal G}_d\\
 0, & i\in{\cal G}_0\setminus({\cal G}_g\cup{\cal G}_d)\end{array}\Biggr.
 \label{eqn:flow_cond}\\
 && \forall \{i,j\}\in{\cal G}_1:\ \ \theta_i-\theta_j=x_{ij}p_{ij}
 \label{eqn:DC_cond}
\end{eqnarray}
where $x=(x_{ij}|\{i,j\}\in{\cal G}_1)$, $g=(g_i|i\in {\cal G}_g)$, $d=(d_i|i\in{\cal G}_d)$, $\theta=(\theta_i|i\in{\cal G}_0)$, $p=(p_{ij}=-p_{ji}|\{i,j\}\in{\cal G}_1)$ are the vector of line inductances, the vector of powers injected at generators, the vector of demands consumed at loads, the vector of phases and the vector of line flows, respectively. (Here $\{i,j\}$ is our notation for directed edges and $j\sim i$ indicates that $j$ is the graph neighbor of $i$.) Note that to streamline notations, we used an abbreviated version of the DC power flow equations in (\ref{eqn:flow_cond},\ref{eqn:DC_cond}). In particular, we ignore terms associated with tap transformers. In our simulations we utilize the DC-PF solver from the Matlab based MATPOWER package \cite{ZimSanTho2010} taking into account effects of transformers and other devices included in the description of the Polish grid model.

\subsection{Optimal Power Flow}
\label{subsec:standard_dc_optimal_power_flow}

Our base solution is obtained by solving the standard DC optimal power-flow problem, finding the optimum generator dispatch given the initial load $d^0$ and cost functions $f=(f_i|i\in {\cal G}_g)$ for every generator as well as generation power and line capacity constraints. To execute this task we use MATPOWER \cite{ZimSanTho2010}, and cost functions provided in the description of the Polish model. The DC optimal power-flow, in the simplest nomenclature, corresponds then to solving
\begin{eqnarray}
\left.\min_{p, g, \theta} \sum_i f_i(g_i)\right|_{
\begin{array}{c}
\mbox{Eqs.~(\ref{eqn:flow_cond},\ref{eqn:DC_cond}), where $d\to d^{0}$}\\
\forall \{i,j\}:\quad |p_{ij}|\leq p_{ij}^{\max}\\
\forall i:\quad g_i^{\min} \leq g_i \leq g_i^{\max}
\end{array}}
\label{OPF}
\end{eqnarray}
for the branch flows, $p$, and generation powers, $g$. The resulting $p^0$, $g^0$ and $\theta^0$ form  the base (reference) solution for our cascading algorithm.
\subsection{Identify islands}
\label{subsec:check_for_islanding}

Our algorithm does not generate a surviving balanced sub-grid at once, but instead resolves it in steps, mimicking dynamics of realistic cascades. The temporal evolution of the surviving sub-grid  is induced by cutting saturated lines, which might also cause the formation of islands and removing freshly formed but overloaded islands.  We check for islanding (i.e. splitting of the grid into independent components) using a depth-first-search based algorithm. If an island is formed, we do all other computations within the cascading algorithm independently for every island.

\subsection{Droop Control and Load Shedding}
\label{subsec:droop_control}

In the process of evaluating the cascading algorithm it can happen, due to tripping of overloaded lines, that some loads or generators will become disconnected from the grid or that the grid splits up into islands. Both scenarios require fast automatic redistribution of generation, done in the so-called droop control fashion~\cite{Kundur1994}.

Droop control is executed at each generator locally in response to an increase or decrease of the system frequency (measured locally as well). Droop control in our algorithm is necessary if the grid changes its structure, i.e. following the appearance of new island(s) in the result of line tripping.
Here we assume that the power generation, $g_i(+)$, at node $i$ after this events is
\begin{equation}
g_i(+)=\frac{g_i(-)}{g_\Sigma(-)}d_\Sigma(+),
\label{eqn:droop_control}
\end{equation}
where the newly introduced quantities on the right hand side of Eq.~(\ref{eqn:droop_control}) are the
current power generation, $g_i(-)$, at node $i$;
the total power generation (before droop control), $g_\Sigma(-)=\sum_{j\in \Sigma_g} g_j(-)$, at the freshly formed island, $\Sigma\subset{\cal G}$, the generator belongs to;
and the total power demand, $d_\Sigma(+)=\sum_{j\in\Sigma_d} d_j(+)$, of the island.
Droop control is executed at every generator of the grid instantaneously. Note that the ratio on the right hand side of eq.~(\ref{eqn:droop_control}) changes in the process of our discrete event simulations in accordance with the modification of islands. If at some point in the process a generator becomes saturated, we do not include it anymore in the droop control mechanism described above, but instead keep its generation level constant (at the maximum generation capacity). As long as demand and total power generation can be matched, the island persists. If the total demand in the island exceeds generation capacity, one first tries to shed $10\%$ of the loads,  and then shuts the island down only if the latter is not successful in balancing the remaining load. This load shedding scheme is a simple proxy of the ``real world" mandatory load shedding implemented in cases of extreme emergency. To inspect the effect of load shedding, we also compare the results against an even simpler strategy considered in \cite{PfiTurChe2011}. Here we skip the load shedding step and switch off an overloaded island immediately.

\section{Random (Natural) Tripping}
\label{sec:Random}

\begin{figure}[b]
\includegraphics[width=1\columnwidth]{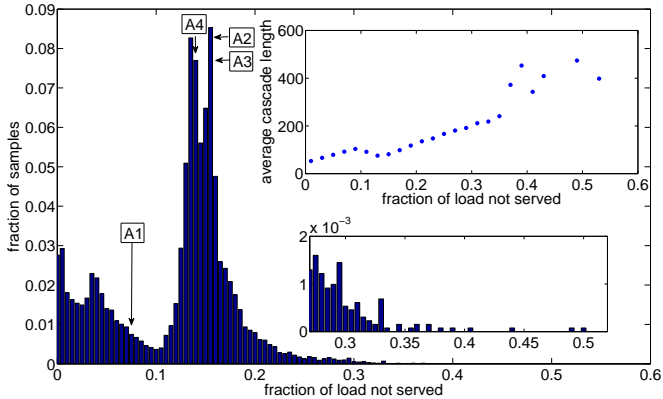}
\caption{Histogram of different outage sizes of 13.000 samples, initiated by tripping line \textit{44}. This line is from the top $1\%$ of the most stressed lines (graded in power flows). Every instance was initiated identically. Bottom inset zooms in the largest values. Top inset shows the average length of cascades, measured in the number of sequentially tripped lines.}
\label{fig:2736sp_tripline44_20000samples}
\end{figure}

\begin{figure}%[t]
\includegraphics[width=1\columnwidth]{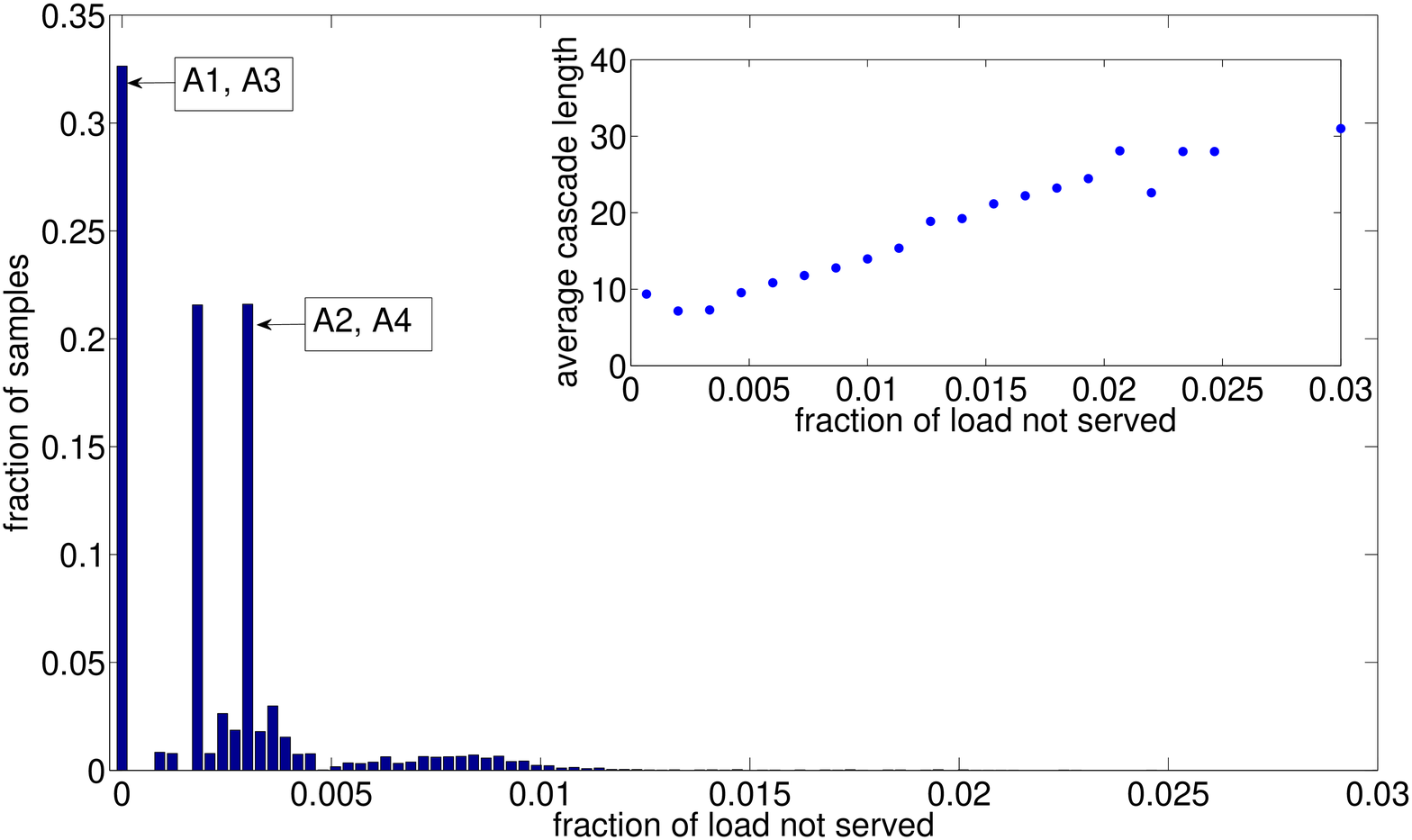}
\caption{Histogram of different outage sizes of 20.000 samples, initiated tripping line \textit{2832}.
Notations and descriptions of the insets are as in the caption to Fig.~(\ref{fig:2736sp_tripline44_20000samples}).}
\label{fig:trip_line_2832}
\end{figure}

\begin{figure}%[t]
\includegraphics[width=1\columnwidth]{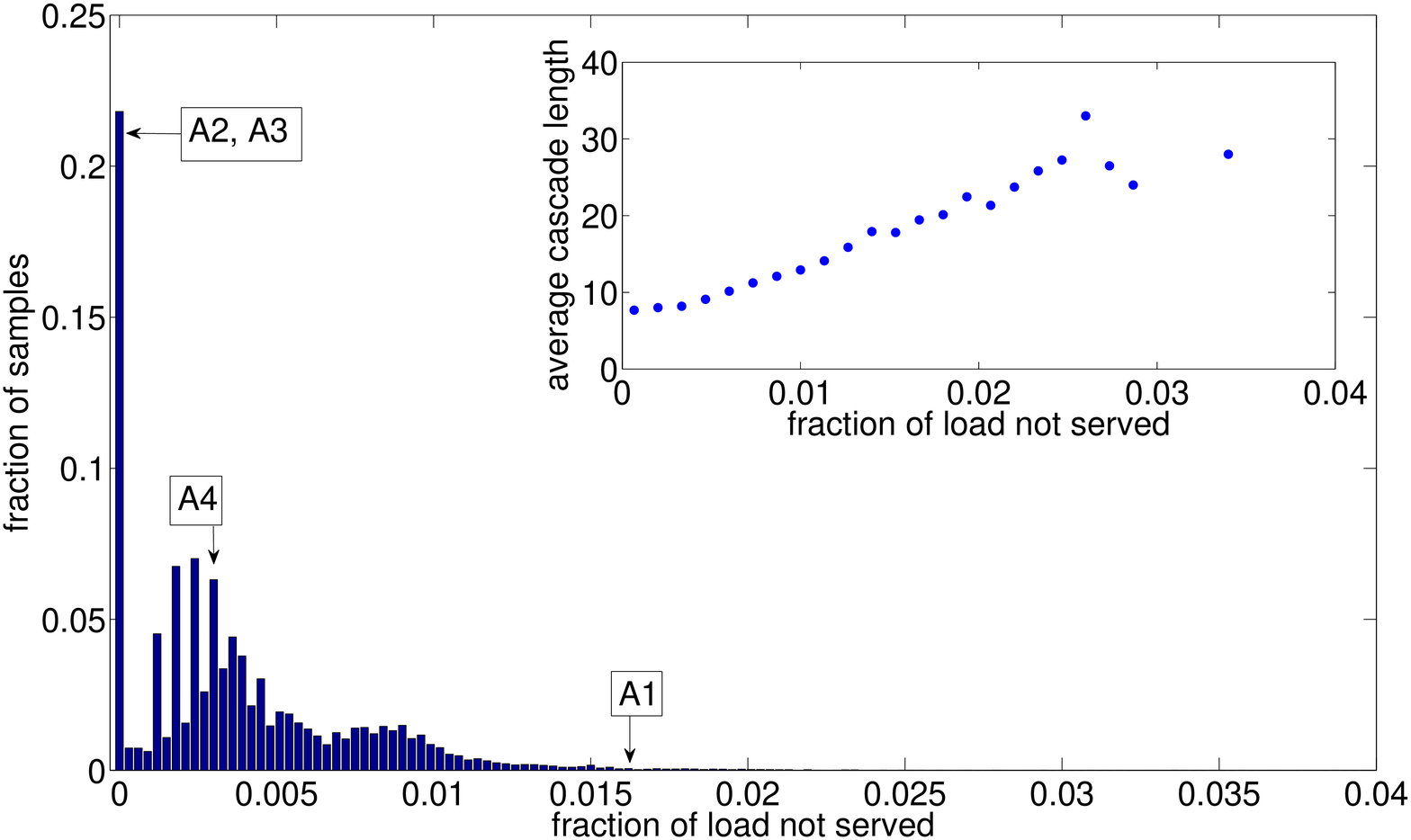}
\caption{Histogram of different outage sizes of 20.000 samples, initiated by tripping line \textit{102}.
Notations and descriptions of the insets are as in the caption to Fig.~(\ref{fig:2736sp_tripline44_20000samples}).}
\label{fig:trip_line_102}
\end{figure}

\begin{figure}%[t]
\includegraphics[width=1\columnwidth]{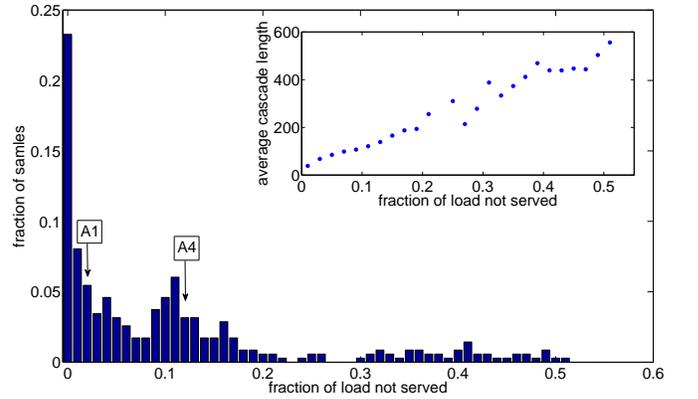}
\caption{Histogram of different outage sizes of 350 samples, initiated by tripping lines \textit{3} and \textit{29}.
Notations and descriptions of the insets are as in the caption to Fig.~(\ref{fig:2736sp_tripline44_20000samples}).}
\label{fig:trip_line_3_29}
\end{figure}

We first consider random (natural) tripping.  The idea is straightforward: at each step of the cascade consider all overheated lines on equal footing and pick one of the lines at random. This uniform and memoryless direct sampling generates a tripping path which stops eventually.  We repeat this process many times for each initial tripping. The results are presented in the form of histograms in Figs.~(\ref{fig:2736sp_tripline44_20000samples},\ref{fig:trip_line_2832},\ref{fig:trip_line_102},\ref{fig:trip_line_3_29}) for four representative examples of initial tripping(s). Of the four examples considered,  the first three shown in Figs.~(\ref{fig:2736sp_tripline44_20000samples},\ref{fig:trip_line_2832},\ref{fig:trip_line_102}) correspond to initial tripping of a single line, enumerated as line \textit{44}, \textit{2832} and \textit{102} respectively. In the last example shown in Fig.~(\ref{fig:trip_line_3_29}) we initiate by tripping two lines simultaneously (line \textit{3} and \textit{29}).

We motivate our choice of initial tripping by the following considerations:
\begin{itemize}
\item Tripping a single line (or a small number of not highly loaded lines) will most likely not lead to any cascade at all, due to the $n-1$ contingency criterium. Hence, we created the candidate list of the top 1\% most loaded lines.

\item In pre-simulations (data not shown) we sampled over all possible initial trippings of single lines within this list and studied the resulting blackout size. Most of the pre-samples did not produce any notable outage, what we attribute to enforcement of the $n-1$ contingency constraint in the base case. Out of the ``bad" samples, producing notable outage, we chose lines \textit{44}, \textit{2832} and \textit{102} for the simulations discussed in the manuscript.

\item Contingency tripping of more lines gets unlikely in a real scenario. However, we choose again to do pre-sampling over all combinations of initially tripping exactly two lines out of the 1\%-list and studied the resulting blackout size. Again, most combinations (not containing any of three in the previous bullet chosen lines) did not result in significant outage. Out of the ones which did, we chose the combination of tripping line \textit{3} and \textit{29} for our main simulations.
\end{itemize}

There are a number of important observations one can make from analyzing the histograms shown. First of all, the four examples (with different initiations) are all different in terms of the average size of the outage,  even though they all correspond to approximately the same amount of the initial power loss. Indeed,  in the example with tripping line \textit{44} and the example where  lines \textit{3} and \textit{29} were tripped simultaneously, the average damage is significantly larger than in the other two examples considered. Second,  and probably most importantly, in all of the four cases the resulting distribution is rather broad. Given the severity and costs of a large blackout, it is especially troublesome to observe that in all these examples ending in a serious outage is very much possible. This suggests that choosing a particular sequence of trippings can make a big difference. This observation motivated the development of relatively simple (on-line and memoryless) control heuristics discussed in the next Section. Third,  in three of the four cases the number of good instances (with zero or small final damage) is significant and even in the worst case (tripping of line \textit{44}) it is still nonzero. This suggests that if one can run fast reliable simulations off-line, simply sampling the tripping paths uniformly and in a memoryless fashion, then picking the sequence giving the least final damage and tripping the actual lines (on line) accordingly, would give a very reasonable control scheme. (Note that this sampling strategy will only suffice if the state of the system is well known. If computational resources and the time allocated for the off-line computations are sufficient the sampling approach will work well. On the other hand, when the tripping action needs to be taken care of immediately one will need to rely on a more efficient control heuristic.) Finally, we observe that the histograms are not monotonic, often showing second and sometimes third local maxima.  We associate this observation with complexity 
of the underlying network and microscopic resolution of cascade process.

The insets of Figs.~(\ref{fig:2736sp_tripline44_20000samples},\ref{fig:trip_line_2832},\ref{fig:trip_line_102},\ref{fig:trip_line_3_29}) also show scatter plots relating the length of the cascade (measured in terms of the number of lines tripped) to the size of the resulting blackout (fraction of load not served).  The main observation here is that the two characteristics are strongly correlated: the larger the outage the longer the cascade.

\section{Effect of load-shedding}
\label{sec:LoadShedding}

\begin{figure}%[b]
\includegraphics[width=1\columnwidth]{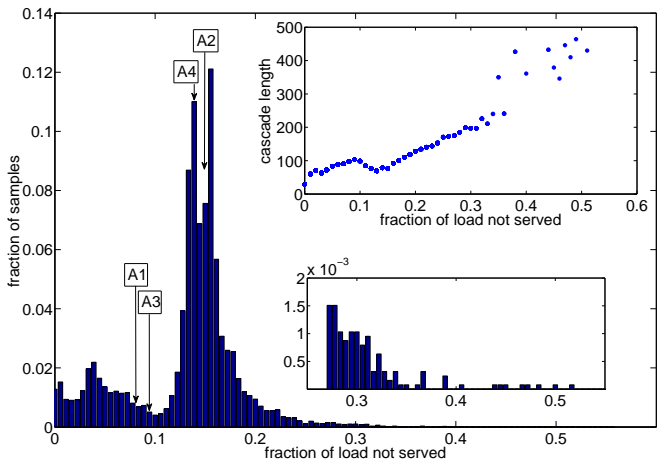}
\caption{The no-load-shedding version of the histogram shown  in the Fig.~\ref{fig:2736sp_tripline44_20000samples}.}
\label{fig:2736sp_tripline44_20000samples_old}
\end{figure}

Fig.~\ref{fig:2736sp_tripline44_20000samples_old} shows results equivalent to these shown in Fig.~\ref{fig:2736sp_tripline44_20000samples}, however \textit{excluding} load-shedding and thus switching off an overloaded island immediately. We discover  that the overall effect of the load shedding does not lead to any significant qualitative changes in the outage distribution function, even though some quantitative changes were observed, in particular in the low-outage scenarios which show a slightly better performance in the case without load shedding.

\section{Control of the Tripping Path}
\label{sec:Control}

The aforementioned observation of the strong sensitivity of the outage size to changes in the tripping sequence suggests that designing the optimal strategy,  leading to the smallest outage, is an extremely important problem.  It is also a hard problem, due to the complex topology and dynamics of the network. We will not attempt to approach this difficult problem systematically here.  Instead,  we suggest to test some simple memoryless heuristics,  i.e. algorithms choosing the next-to-be-tripped line based on the current state of the active part of the grid. We tested the following four schemes:
\begin{enumerate}
\item [\textbf{A1}] Trip the line, $(i,j)$, with the minimal current power flow, $P_{ij}=\min\{P_{\cal O}\}$;
\item [\textbf{A2}] Trip the line, $(i,j)$,  with the maximal current power flow, $P_{ij}=\max\{P_{\cal O}\}$;
\item [\textbf{A3}] Trip the line, $(i,j)$, with the minimal current relative overload, $p_{ij}=\min\{p_{\cal O}\}$;
\item [\textbf{A4}] Trip line, $(i,j)$, with the maximal current relative overload, $p_{ij}=\max\{p_{\cal O}\}$;
\end{enumerate}
where the relative overload is defined as $p_{ij}=(P_{ij}-P^{\text{max}}_{ij})/P^\text{max}_{ij}$.

Our choice of the four strategies is motivated by the following considerations.\\
\textbf{(A1):}
Strategy (A1) was inspired by considering an oversimplified hierarchical model with a tree-like structure, the top level representing the transmission part of the grid. The lower in the hierarchy the line is positioned, the more likely it is to carry less power. In this picture the majority of generators would be on a high level in the hierarchy. Tripping an overloaded line of comparably small power flow will correspond to tripping a line at a low level, thus leading to cutting a small sub-tree. This sub-tree possibly contains fewer generators than needed to support their demands. Hence, the sub-tree will most likely shut down. Since the tripped line is (most probably) from the low level, the shut down demand will be rather small. This kind of ``load shedding'' will effectively de-stress the system and eventually lead to a stable power flow solution with no overloaded lines on higher levels. Obviously the Polish grid is not cleanly hierarchical,  containing sufficient number of loops. However,  the number of edges and vertexes in this relatively large grid is roughly the same, suggesting that at least locally the graph is somewhat tree like. This observation is also supported by Fig.~\ref{fig:polish1}, showing an illustration of the topology of the Polish power grid. Overall,  the above arguments suggest that the tree/hierarchy based strategy (A1) may be working reasonably well.\\
\textbf{(A2):}
The second strategy  was suggested as the inverse of (A1). It is also a greedy strategy: taking care of the worst local problem, and thus ignoring any possible long correlations and structural connections.\\
\textbf{(A3):}
The logic behind (A3) is similar to (A1),  however suggesting to use the relative overload instead of the absolute overload as an alternative (and possibly more accurate) measure of the line stress. And indeed, the (A3) scheme shows performance very similar to (A1) in our base experiments.  However, it becomes less efficient (than (A1)) in the simplified trials, ignoring mandatory load shedding.
\\
\textbf{(A4):} This strategy is similar to (A2) in what concerns being greedy and focusing first on the worst local problem. The difference with (A2) is in the replacement of the absolute overload criterium of (A2) by the relative (and re-scaled) one in (A4).  One may argue that the relative criterium of (A4) mimics the ``natural'' sequence of trippings (i.e. the one which takes place  without ``line switching'' control) better than (A2) and better than tripping overloaded lines uniformly. Indeed, the probability of natural tripping of a line does not depend on other lines (therefore, it is local), and it will also be a fast (possibly exponential) growing function of the line relative overload.
We want to stress that all these decision schemes relay on fast uplink communications (to a central location) of the current line status (overloaded or not,  and if overloaded by how much) followed by a tripping signal communicated downlink.

We compare performance of the four schemes against each other and also against the random uniform trippings in Figs.~(\ref{fig:2736sp_tripline44_20000samples},\ref{fig:trip_line_2832},\ref{fig:trip_line_102},\ref{fig:trip_line_3_29}). We observe that our hierarchical interpretation of the Polish grid was in majority of cases (but not always) reasonable.  In all, but the third example of Fig.~(\ref{fig:trip_line_102}), (A1) is a clear cut winner, and generally (A2) and (A4) are performing worse than (A1) and (A3). Also, and quite remarkably, uniform tripping was rather successful in the cases of Figs. \ref{fig:trip_line_2832},\ref{fig:trip_line_102} and \ref{fig:trip_line_3_29}, resulting in no damage (or almost no damage) for significant, $O(1)$, number of samples. In the case of the Fig. \ref{fig:trip_line_3_29}, random tripping leads to almost no damage in $22\%$ of samples, note however that the worst of the $400$ samples has lead to a very severe damage - removal of $60\%$ of loads. In general, the tests were inconclusive in terms of looking for a universally reliable heuristics. It leads us to believe that finding the optimal strategy is not going to be easy,  and hence making massive off-line sampling (given reliable state estimations) will likely be the most reliable choice.

\section{Summary}
\label{sec:Summary}

The research presented in this manuscript extends and complements previous studies of control and optimization for mitigating blackouts and vulnerability analysis of power grids \cite{BieMat2007, Bienstock2010, ZdeDecChe2009, PinMezDonLes2010, HinTal2007}. Of many possible theoretical control actions capable to mediate emerging cascade, such as generator dispatch or load shedding, we focus here solely on analysis of \textit{tripping of already overloaded lines}. In what concerns sequential removal of overloaded lines, our numerical study can also be viewed as suggesting dynamic extension of the static $N-k$ contingency problem analyzed in \cite{BieVer2010}. We perform our study on a real-world power grid structure, using a microscopic DC power flow approximation, however to generalize our analysis to the general AC framework will be straightforward. We showed that controlled tripping of overloaded lines may lead to significant mitigation of the resulting damage, as it forces the cascade to go through a less damaging scenario as if it would develop by itself without the mitigation. The problem of finding an universally optimal sequence of trippings is computationally hard. To mitigate the hardness we settle in this study on suggesting and analyzing some plausible tripping heuristics, formulated as graph-local searches over the current state of the grid, which is memoryless and requires efficient communications of the SCADA type. Of the strategies considered,  the heuristics performing the best in the majority of cases suggest to trip the least overloaded line first. We plan to extend this study by analyzing other grids, work on improving algorithm,  e.g. analyzing more sophisticated optimization schemes including searching through strategies over anticipated future moves (with time horizon), and analyze hybrid control schemes combining line tripping with other actions, such as emergency generation dispatch.

\section{Acknowledgements}

We are thankful to the participants of the ``Optimization and Control for Smart Grids" LDRD DR project at Los Alamos and Smart Grid Seminar Series at CNLS/LANL for multiple fruitful discussions. We also like to acknowledge stimulating discussions with Prof. M. Kezunovic, who brought the idea of already overloaded line tripping, as a viable option for emergency control, to our attention.

\ifCLASSOPTIONcaptionsoff
%  \newpage
\fi

\bibliographystyle{IEEEtran}
\bibliography{SmartGridComm}

\vspace{-1.0cm}
\begin{IEEEbiographynophoto}{Ren\'{e} Pfitzner}
is a PhD student at the Chair of Systems Design, ETH Z\"{u}rich, Switzerland. Previously he worked with M. Chertkov at Los Alamos National Lab and New Mexico Consortium, Los Alamos, NM, USA. {\tt\small rpfitzner@ethz.ch}
\end{IEEEbiographynophoto}
\vspace{-1.0cm}
\begin{IEEEbiographynophoto}{Konstantin Turitsyn}
received his PhD in Theoretical Physics in 2007 from Landau Institute, Moscow. He was a postdoctoral
scholar at University of Chicago and a J. R. Oppenheimer Fellow at Los Alamos National Lab. He is now an Assistant Professor at MIT, Department of Mechanical Engineering, MA, USA. {\tt\small turitsyn@mit.edu}
\end{IEEEbiographynophoto}
\vspace{-1.0cm}
\begin{IEEEbiographynophoto}{Michael Chertkov}
received his PhD in physics from the Weizmann Institute of Science in 1996. He is a staff-member and PI for the ``Optimization and Control Theory for Smart Grids" project at Los Alamos National Lab. {\tt\small chertkov@lanl.gov}\end{IEEEbiographynophoto}

\end{document}